\documentstyle[12pt,preprint,aps]{revtex}

\draft

\newcommand{\be}{\begin{eqnarray}}
\newcommand{\ee}{\end{eqnarray}}

\preprint{\normalsize NUC-MINN-97/11-T}

\title{NUCLEUS-NUCLEUS BREMSSTRAHLUNG FROM ULTRARELATIVISTIC COLLISIONS}
\author{
Sangyong Jeon\footnote{jeon@nucth1.spa.umn.edu}
and Joseph Kapusta\footnote{kapusta@physics.spa.umn.edu}}
\address{
School of Physics and Astronomy\\
University of Minnesota\\
Minneapolis, MN 55455}
\author{
Alexei Chikanian\footnote{chikanian@hepmail.physics.yale.edu}
and Jack Sandweiss\footnote{sansweiss@hepmail.physics.yale.edu}}
\address{
Department of Physics\\
Yale University\\
New Haven, CT 06520}

\begin{document}

\maketitle
\thispagestyle{empty}

\begin{abstract}
The bremsstrahlung produced when heavy nuclei collide is estimated
for central collisions at the Relativistic Heavy Ion Collider.
Soft photons can be used to infer the rapidity distribution
of the outgoing charge.  An experimental design is outlined.
\end{abstract}

\setcounter{page}{1}
\section{Introduction}

The Relativistic Heavy Ion Collider (RHIC) at Brookhaven National
Laboratory (BNL) will have its first beam available for experiments
by the end of 1999.  A basic issue is how transparent the nuclei
are to each other.  This issue is important because it determines
the initial energy density of hot matter whose study is, after all,
the ultimate physics goal.  The definition of transparency is not
unique; transparency may be realized as the deceleration of baryons,
or of electric charge, or the number of hadrons produced.  The fact
that these different realizations are not tightly coupled may be
illustrated by the recognition that in a cascade picture the number
of produced particles decreases each time a baryon collides because
of the decreased energy available in comparison with the previous
collision.  Hence the first baryon collision is the most significant
from the particle production point of view and the last is the
least significant.  In contrast, each collision slows the baryon
and each is equally significant because the rapidity loss per
collision is practically independent of beam energy.  In this
paper we shall focus on the deceleration of charge as a global
indicator of transparency.  The advantage of charge over baryon
number has to do with experimental detection.  A significant
fraction of baryon number may emerge in the form of neutrons
and hyperons whose detection is complicated.  Global stopping
of electric charge is reflected in the amount of bremsstrahlung
which can be readily computed and detected; this is the focus
of the paper.
Of course, the net charge as a function of
rapidity can also be measured directly, but this requires
identification of all charged particles over all of momentum
space, and is much more difficult than the experiment to
be discussed here.

The first theoretical study of nucleus-nucleus bremsstrahlung
was done by one of the authors \cite{kap} and intended for
application at the Bevalac where the beam energy was in the
range of 250 to 2100 MeV per nucleon in the laboratory frame.
A later study was done by Bjorken and McLerran \cite{BM} which
was aimed at much higher energies.  At the time of that study
it was anticipated that the central rapidity region would be
relatively free of baryons because of high transparency \cite{Bj}.
Recent experimental studies at the SPS at CERN \cite{expt}
are suggesting
that many more baryons than anticipated will emerge at central
rapidities.  The baryon and charge rapidity distributions at
RHIC may even be flat!  Therefore it makes sense to reinvestigate
the bremsstrahlung to be expected from central collisions of the
heaviest beam nuclei at RHIC\footnote{The study by Bjorken and McLerran
was directed towards fixed target experiments.  As an aside they boosted
the bremstrahlung distribution to the center of momentum frame, as
befits a collider, but made a sign error in the Lorentz transformation.
This led to the expectation that the bremmstrahlung would be in
the GeV range, whereas it ought to be in the MeV range, as shown below.}
and a potential experimental design to measure it.

Emission of soft photons was also explored theoretically by
Dumitru, McLerran, Stocker and Greiner \cite{Dumitru} in the
context of the RQMD (Relativistic Quantum Molecular Dynamics)
model.  They point out that $\pi^0$ decay photons contribute
a negligible amount to the soft photon spectrum at forward angles,
suggesting the feasibility of the bremsstrahlung measurement.  They
propose to study charge stopping by measuring the total
energy radiated as soft photons.  Since the spectrum
peaks at a very small angle, the measurement of total energy
is actually impossible at RHIC because the detector would
have to be placed within the beam pipe and would interfere with
the operation of the collider.  We will show that measurement
of the spectrum at several angles in the range from about 2
to 10 degrees is sufficient to discriminate different stopping
scenarios, and outline a detector design to do so.

\section{Calculation of Bremsstrahlung}

Consider a central collision of two equal mass nuclei of charge $Z$
in their center of momentum frame.  Denote the speed of each beam by
$v_0$ and the corresponding rapidity by y$_0$, so that the
rapidity gap between projectile and target nuclei is 2y$_0$.
Velocity and rapidity are related by $v = \tanh {\rm y}$.
Before collision each nucleus may be viewed as a flattened pancake
with negligible thickness.  The current carried by the projectile
is:
\begin{equation}
{\bf J}_{\rm P}({\bf x},t) = v_0 \sigma(r_{\perp})
\delta(z-v_0t) \theta(-t) {\bf \hat{z}} \, ,
\end{equation}
and by the target is:
\begin{equation}
{\bf J}_{\rm T}({\bf x},t) = - v_0 \sigma(r_{\perp})
\delta(z+v_0t) \theta(-t) {\bf \hat{z}} \, ,
\end{equation}
where $\sigma(r_{\perp})$ is the charge per unit area at a distance
$r_{\perp}$ from the central beam axis.
For low frequency photons the nucleus-nucleus collision appears
almost instantaneous.  To the extent that the transverse rapidities
of the outgoing charged particles are small compared to the beam
rapidity one may then approximate the current after the collision as:
\begin{equation}
{\bf J}_{\rm F}({\bf x},t) = {\bf \hat{z}} \sigma(r_{\perp}) \theta(t)
\int_{-\infty}^{\infty} d{\rm y} \rho(r_{\perp},{\rm y}) v({\rm y})
\delta(z-v({\rm y})t) \, .
\end{equation}
Here $\rho$ represents the charge rapidity distribution and is normalized as:
\begin{equation}
\int_{-\infty}^{\infty} d{\rm y} \rho(r_{\perp},{\rm y}) = 2 \, ,
\end{equation}
the 2 arising because the total charge is 2$Z$.

The classical amplitude to emit electromagnetic radiation
in the direction ${\bf n}$ with frequency $\omega$ is \cite{Jackson}:
\begin{equation}
{\bf A}({\bf n},\omega) = \int dt \int d^3x \, {\bf n}\times\left(
{\bf n}\times {\bf J}({\bf x},t)\right) e^{i\omega(t-{\bf n}
\cdot {\bf x})} \, .
\end{equation}
Here ${\bf J} = {\bf J}_{\rm P} + {\bf J}_{\rm T} + {\bf J}_{\rm F}$
is the total current.  It is convenient to take ${\bf n} =
\left(\sin\theta, 0, \cos\theta \right)$.  Then the distribution
in frequency and direction is:
\begin{equation}
\frac{d^2N}{d\omega d\Omega} = \frac{\alpha}{4\pi^2 \omega}
\sin^2\theta \left| \int dx dy \sigma(r_{\perp}) e^{-i x \omega \sin\theta}
\left[ \int d{\rm y} \frac{v({\rm y}) \rho(r_{\perp},{\rm y})}
{1-v({\rm y})\cos\theta} - \frac{2v_0^2 \cos\theta}{1-v_0^2
\cos^2\theta} \right] \right| ^2 \, .
\end{equation}
We cannot go further without some knowledge of the distribution
$\rho(r_{\perp},{\rm y})$.

In Fig. 1 we plot the quantity $\rho(r_{\perp},{\rm y})$
as computed in LEXUS \cite{lexus} for central Au+Au collisions
at 100 GeV per nucleon in the cm frame.
LEXUS is a linear extrapolation of nucleon-nucleon
scattering to nucleus-nucleus collisions.  It is based on sequential
nucleon-nucleon scatterings, as in free space, with energy loss taken
into account.  For $r_{\perp}=0$ this distribution has a broad maximum
at y = 0, whereas for $r_{\perp} \approx R$, the nuclear radius, this
distribution has a broad minimum at y = 0.  Overall the charge rapidity
distribution is roughly flat.  We have computed the bremsstrahlung
from LEXUS and will display it shortly.

It is also advantageous to have a simple analytic model with a
variable charge rapidity distribution.  To this end suppose that
$\rho(r_{\perp},{\rm y})$ is independent of $r_{\perp}$.  Then
\begin{equation}
\frac{d^2N}{d\omega d\Omega} = \frac{\alpha Z^2}{4\pi^2 \omega}
\sin^2\theta \left| F(\omega \sin\theta) \right|^2 \left|
\left[ \int d{\rm y} \frac{v({\rm y}) \rho({\rm y})}
{1-v({\rm y})\cos\theta} - \frac{2v_0^2 \cos\theta}{1-v_0^2
\cos^2\theta} \right] \right| ^2 \, ,
\end{equation}
where $F$ is a nuclear form factor:
\begin{equation}
F = \frac{1}{Z} \int dx dy \, \sigma\left( \sqrt{x^2+y^2} \right)
e^{-i x \omega \sin\theta} \, .
\end{equation}
A solid sphere approximation should be adequate for large nuclei,
in which case
\begin{equation}
F(q) = \frac{3}{q^2}\left( \frac{\sin q}{q} - \cos q \right) \, ,
\end{equation}
where $q=\omega R \sin\theta$ and $R$ is the nuclear radius.
Actually, for the range of angles and frequencies of interest to us,
the nuclear form factor is practically equal to one.

The integral over rapidity can easily be performed for
a flat rapidity distribution.
\begin{equation}
\rho({\rm y}) = \frac{\theta({\rm y}_0-{\rm y}) \theta({\rm y}_0+{\rm y})}
{{\rm y}_0}
\end{equation}
Then the photon distribution is:
\begin{eqnarray}
\frac{d^2N}{d\omega d\Omega} &=& \frac{\alpha Z^2}{4\pi^2 \omega}
\sin^2 \theta \left| F(\omega R \sin\theta) \right|^2 \nonumber \\
&\times& \left[ \frac{2}{\sin^2 \theta} -
\frac{1}{{\rm y}_0 \sin^2 \theta}
\ln\left( \frac{1+v_0 \cos\theta}
{1-v_0 \cos\theta} \right) - \frac{2v_0^2 \cos\theta}{1-v_0^2
\cos^2\theta} \right] ^2 \, .
\end{eqnarray}
Because $v_0$ is close to 1 the distribution is highly modifed from
the quadrupole form, being strongly peaked near small but nonzero
$\theta$.  At RHIC the peak occurs at $\theta \sim 1^{\circ}$.

Since the frequency dependence of bremsstrahlung photons is $1/\omega$,
for low frequencies anyway, there is no loss in information in integrating
over a range of frequencies.  For the detector described in the
next section we can consider all photons with energies between
10 keV and 3 MeV.  Fig. 2 shows the spectra for central
gold collisions at RHIC using both LEXUS and a flat charge rapidity
distribution which is independent of $r_{\perp}$.  The two models
give almost identical results, as may be expected based on Fig. 1.

We have also computed the spectra resulting from
three different final charged particle rapidity shapes that
span the possibilities at RHIC.
These shapes are shown in  Fig. 3.  The curve $\alpha$ represents a
maximal stopping scenario, $\beta$ an approximately 50\% stopping case, and
$\gamma$ the case of minimal stopping.
The results are shown in Fig. 4.
Once again we have taken only photon energies between 10
keV and 3 MeV so as to match the experimental design described in the
following section.  There is a very clear difference in the photon
spectrum between the cases of ``full stopping" and "transparency".
The cases of ``50\% stopping" and a flat rapidity distribution
are very similar, although they are still very differenct from the
other two cases, and even these can be distinquished with sufficient
statistics and angular resolution in an experiment.

To conclude this section we note that the classical bremsstrahlung
formula ought to be accurate for photon energies less than
$\hbar /\tau$, where $\tau$ is the time elapsed between first
nuclear contact and the last hard scattering of the charged hadrons.
For $\tau$ on the order of 10 to 50 fm/c this restricts the validity
to photons with energies less than 20 to 4 MeV, respectively.

\section{A Detector Design}

The bremsstrahlung yields calculated above are measureable in a relatively
modest experiment. The basic features of the bremsstrahlung which allow this
are the low energies, less than a few MeV,
and concentration to forward angles, with almost all of the
radiation within $10^{\circ}$ of the beam line.
In the following we describe a simple experiment which could detect
and measure this bremsstrahlung. This is meant as an example of how it might be
done. Of course, a real experiment would need to be concerned with many other
details and might well choose somewhat different dimensions, number of
detectors, and so on. The discussion presented below is limited in detail
as appropriate for this paper. However, we believe that sufficient information
is presented to show the feasibility of the measurement.

\subsection{Direct $\pi^{0}$ Background}

An experiment to detect and measure this radiation can be based on the detection
of gamma rays with energies ranging from 10 keV to 3 MeV.
At first thought, one might imagine that
 the most dangerous
backgrounds are photons from the decay of $\pi^{0}$ mesons which happen to lie
in this energy range. Because of the large mass of the $\pi^{0}$ (140 MeV) the
typical $\pi^{0}$ decay photon has substantially greater energy. Essentially
the only way a $\pi^{0}$ decay photon can be in this range is for the meson
to be moving away from the detector and then to decay at backward angles, with
just the right kinematics.

As an example, we note that  calculations using HIJET \cite{HIJET} show that
for the experiment described below, the $\pi^{0}$
decay background is smaller than the expected bremsstrahlung rate (for
$3^{\circ}$ and the flat rapidity spectrum) by a factor of $4\times 10^{4}$.
Thus, this background is negligible,
as pointed out already by Dumitru et al. \cite{Dumitru}.
Incidentally, if in the future, very
precise measurements of the bremsstrahlung were desired, this background could
be calculated from the $\pi^{0}$ spectrum, which would be known by that time,
and subtracted.

\subsection{Experimental Configuration}

The photon detector proposed is shown in Fig. 5.
The Ge detector detects the low energy photons by photoabsorbtion for the
low end of the spectrum and Compton scatters for the high end. Detected
energy in excess of 3 MeV will not be accepted as a valid event. The role of
the cylindrical scintillation counter surrounding the Ge cube is to veto higher
energy gamma rays and charged particles. The aluminum cold finger is connected
to a liquid nitrogen supply (not shown) and keeps the Ge detector at liquid
nitrogen temperature.

 The dominant gamma ray flux is from
the decay of $\pi^{0}$ mesons and is in the several GeV range. The pairs
produced by these photons will be detected and vetoed by the scintillation
counter. Compton scatters with final electron energy above about 3 MeV will
also leave the Ge and be vetoed by the scintillator. The precise response of
the counter assembly to the low energies of the bremsstrahlung photons can be
well predicted but can also be calibrated with the use of sources. The detected
energy spectrum will thus not be the true bremsstrahlung spectrum but will be
calculable from the expected bremsstrahlung spectrum (properly weighting
photoabsorbtion and Compton scattering).

Since the shape of the expected frequency spectrum does not
depend on the stopping or the detection angle, the agreement between the
measured and predicted spectral shapes
will be an important check that the experiment is
actually measuring the bremsstrahlung.

One aspect of the detection system is that the background particles not be so
numerous that the apparatus has no live time to detect the real
bremsstrahlung photons. With the collimation scheme proposed below, the live
time is estimated to be 85\%. The magnet and the collimator are necessary to
reduce the dead time due to charged particles to reach this level of live
time.

Figure 6 shows an elevation view of the experiment. (The small counter
system which determines the vertex position is not shown.) The
apparatus has four photon detectors of the type shown in Fig. 5
at polar angles of $3^{\circ}$, $5^{\circ}$, $7^{\circ}$, and $9^{\circ}$.
As can be seen from Figs. 2 and 4,
this permits a characterization of the bremsstrahlung.

A view of the proposed
experiment looking back towards the interaction point from the detector side is
shown in Fig. 7. The collimator slit is narrow and
matched to the detector width. The magnet sweeps charged particles to the left
and right, as seen in Fig. 7. Most of the charged particle flux
then misses the narrow slit in the collimator and does not reach the
detector.  As mentioned above, this feature is necessary to
achieve a sufficiently large live time for the experiment.

\subsection{Backgrounds to the Measurement and Their Solution}

The studies and simulations which have been done indicate that the only
background in the experiment arises from interactions which occur in the beam
pipe, the collimator, the ancillary material, etc. which produce $\pi^{0}$
mesons which in turn produce gamma rays which in turn shower in the
various  materials. The gamma ray showers have a non-negligible probability of
producing a low energy gamma in the bremsstrahlung detection range.
The location of these showers is widely spread out physically and essentially
occurs where the particles produced in the gold-gold collision
interact with the surrounding material.

This background is considerable and is comparable to the bremsstrahlung signal
in rate. Thus, the experiment must measure this background since the
calculations are not precise enough to allow a theoretical subtraction. It
is also essential to use a Be, rather than Al, beam pipe to keep this
background at a
level which can be measured in a reasonable time.

There is a simple way to measure this background. Consider one photon detector.
If a plug, small in area - matched to the small size of the detector - is
placed  behind the collimator,
 between the primary collision and the detector, it will attenuate the
bremsstrahlung flux by a large  amount which can be reliably calculated, and
calibrated with sources.
However, it will
have a negligible effect on the background sources which are present during the
runs without the plug. This is because of the wide distribution of these
sources and the small volume of the plug. The plug has a big effect on the true
signal (attenuates it) because it  completely blocks the path of direct
photons from the interaction to the detector. The plug can be small because the
detector is small, less than 1 cm$^3$ at 4 meters from the interaction.

The situation is
somewhat complicated by the fact that the plug creates new background via the
showering of $\pi^{0}$ gammas in it (which would not have showered in the no
plug run). This means that we have to do the plug experiment twice, with two
different thickness plugs. With the three measurements, one can solve for the
signal rate (and incidentally, of course, the background rates).

An evaluation of the systematics of such a measurement indicates that a
significant measurement could be accomplished in 300 hours of running even at
the initially anticipated RHIC luminosity of 1 central event per second.

\section{Conclusion}

We have shown that the bremsstrahlung emitted during a high energy heavy ion
collision is a sensitive indicator of the degree of stopping of the positive
charges and presumably, therefore, of the transparancy of the collisions.
The bremsstrahlung is emitted at relatively low photon energies,
typically less than a few MeV and
is concentrated at small angles to the line of the colliding beams.

An experimental design has been presented which illustrates that the
bremsstrahlung is measureable at RHIC with a modest apparatus and in a
reasonable time.

\section*{Acknowledgements}
This work was supported by the U.S. Department of Energy under grants
DE-FG02-87ER40328, DE-FG02-90ER40562, and DE-FG02-90ER40704.

\section*{Figure Captions}

\noindent
Figure 1: The distribution $\rho$ as a function of transverse position
and rapidity from LEXUS.\\

\noindent
Figure 2:  The number of bremsstrahlung photons with energies
between 10 keV and 3 MeV as a function of angle for central collisions
of gold at RHIC.  The two curves correspond to LEXUS and a flat charged
particle rapidity distribution.\\

\noindent
Figure 3:  Curves $\alpha$, $\beta$, $\gamma$
represent potential charged particle rapidity spectra for central
collisions of gold at RHIC. They categorize ``full stopping'', ``50\%
stopping'', and ``near transparancy", respectively.
Only the $y>0$ curves are shown:
the spectra are symmetric about y=0.\\

\noindent
Figure 4:  The number of bremsstrahlung photons with energies
between 10 keV and 3 MeV as a function of angle for central collisions
of gold at RHIC.  The three curves correspond to those labelled $\alpha$,
$\beta$, $\gamma$ in the previous figure.\\

\noindent
Figure 5:  Sketch of the bremsstrahlung photon detector.
The small Ge block $(.8 \times .8 \times .5 $ cm$^{3}$)
detects the low energy radiation. The
cylindrical scintillator is operated in anticoincidence to veto pairs and
energetic Compton scatters.\\

\noindent
Figure 6:  Elevation view of the bremsstrahlung experiment.
The magnetic field is along the y axis (vertical) and of strength 20 kG.\\

\noindent
Figure 7:  View of the experiment along the beam,
looking back from the detector side.\\

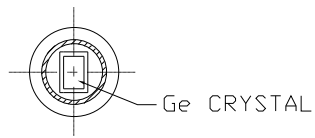
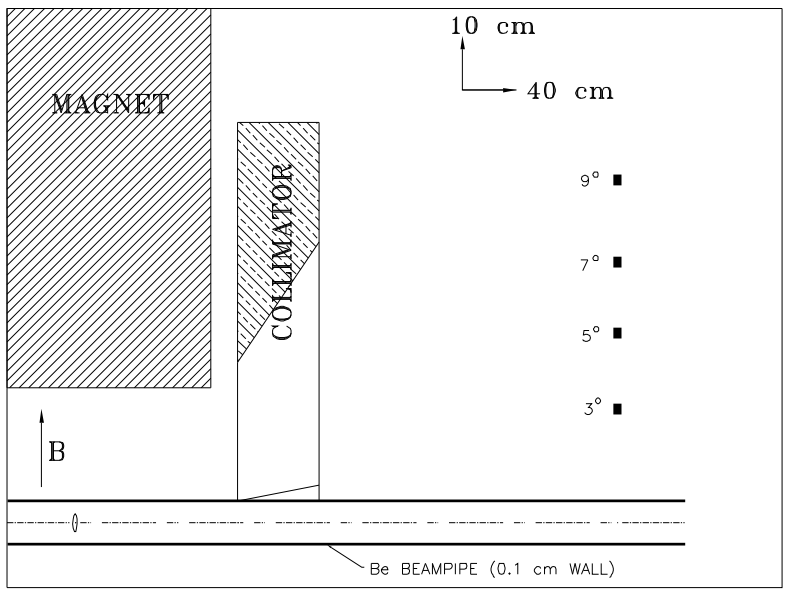
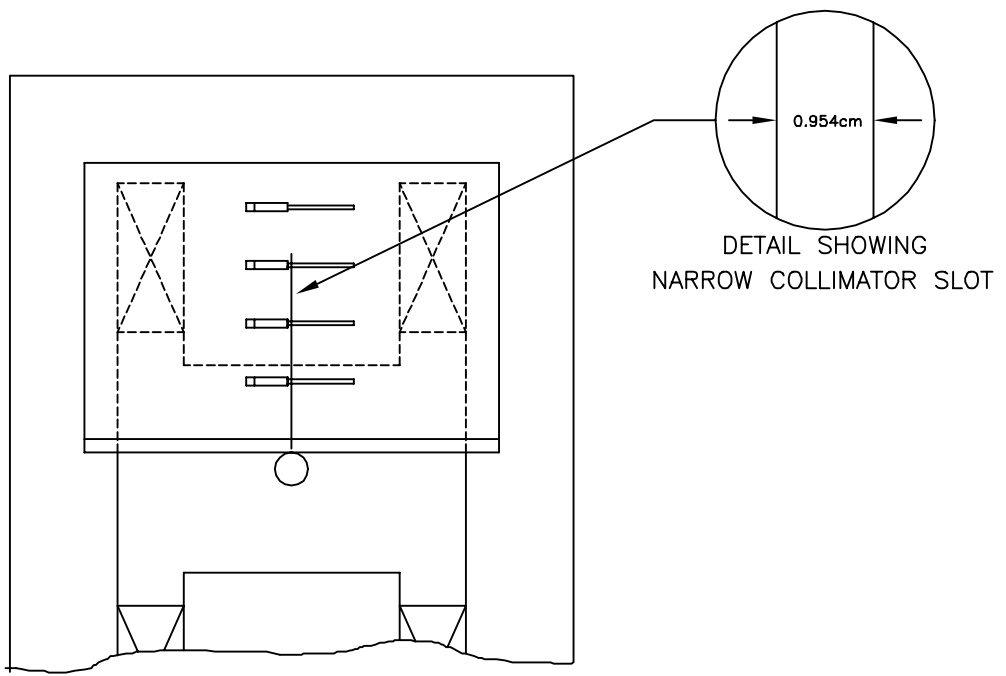
\end{document}